\pdfoutput=1
\documentclass[aps,prb,twocolumn,superscriptaddress,nobibnotes]{revtex4-1}
\usepackage{graphicx}
\usepackage{multirow}
\usepackage{xcolor}
\usepackage{xstring}

\usepackage[normalem]{ulem}
\usepackage{amsmath,amsfonts,amssymb}
\usepackage[colorlinks=true,linkcolor=blue,citecolor=blue,urlcolor=blue]{hyperref}

\newcommand\THEOSMARVEL{Theory and Simulation of Materials (THEOS) and National Centre for Computational Design and Discovery of Novel Materials (MARVEL), {\'E}cole Polytechnique F{\'e}d{\'e}rale de Lausanne, 1015, Switzerland}
\newcommand\Vilnius{Vilnius University Institute of Biotechnology, Saul{\.{e}}tekio al.\ 7, LT-10257 Vilnius, Lithuania}

\graphicspath{{figs/}}

\begin{document}

\title{Two-dimensional materials from  high-throughput computational exfoliation of  experimentally known compounds}
\author{Nicolas Mounet}
\email{nicolas.mounet@epfl.ch}\affiliation{\THEOSMARVEL}
\author{Marco Gibertini}
\affiliation{\THEOSMARVEL}
\author{Philippe Schwaller}
\affiliation{\THEOSMARVEL}
\author{Davide Campi}
\affiliation{\THEOSMARVEL}
\author{Andrius Merkys}
\affiliation{\THEOSMARVEL}\affiliation{\Vilnius}
\author{Antimo Marrazzo}
\affiliation{\THEOSMARVEL}
\author{Thibault Sohier}
\affiliation{\THEOSMARVEL}
\author{Ivano E. Castelli}
\affiliation{\THEOSMARVEL}
\author{Andrea Cepellotti}
\affiliation{\THEOSMARVEL}
\author{Giovanni Pizzi}
\affiliation{\THEOSMARVEL}
\author{Nicola Marzari}
\email{nicola.marzari@epfl.ch}\affiliation{\THEOSMARVEL}

\begin{abstract}
We search for novel two-dimensional materials that can be easily exfoliated from their parent compounds. Starting from 108423 unique,
experimentally known three-dimensional compounds we identify a subset of 5619 that appear layered according
to robust geometric and bonding criteria. High-throughput calculations using van-der-Waals density-functional theory, validated against 
experimental structural data and calculated random-phase-approximation binding energies, allow to
identify 1825 compounds that are either easily or potentially exfoliable, including all that are commonly exfoliated experimentally. In particular, the subset of 1036 easily 
exfoliable cases---layered materials held together mostly by dispersion interactions and with binding energies up to $30-35$ meV$\cdot\text{\AA}^{-2}$---provides a wealth of novel structural prototypes and simple ternary compounds, and
a large portfolio to search materials for optimal properties. For the 258 compounds with up to 6 atoms per primitive cell we comprehensively explore  vibrational, electronic, magnetic, and topological properties, identifying in particular 56 ferromagnetic and antiferromagnetic systems, including half-metals and half-semiconductors.
\end{abstract}

\maketitle
Two-dimensional (2D) materials provide novel opportunities to venture
into largely unexplored regions of materials space.
On one hand, their ultimate thinness makes them extremely
promising for applications in electronics 
(e.g., for field-effect transistors, where
a reduced device size is beneficial to improve performance and
reduce short-channel effects between contacts).\cite{Radisavljevic:2011,Chhowalla2016}
On the other hand, the physical  properties of monolayers often change dramatically from those of parent 3D materials, providing
a new degree of freedom~\cite{butler2013progress} for applications 
(from thermoelectrics to spintronics) while also unveiling novel physics (e.g., valley Hall effect and composite excitations such as trions).
Moreover, van-der-Waals (vdW) heterostructures~\cite{geim2013van}
have recently emerged as an additional avenue to engineer novel properties by stacking 2D materials in a desired fashion.

Still, to date only a few dozens 2D materials have been experimentally synthesised or exfoliated. 
Progress in this area would be strongly accelerated by the 
availability of a much broader portfolio of 2D candidate materials.
To illustrate this point, we can compare the
current situation for known 3D crystals, for which the knowledge accumulated in 
the past century (both crystal structures and measured physical properties)
has been collected in databases such as the Pauling file~\cite{Villars:1998}, 
the Inorganic Crystal Structure Database~\cite{ICSD} (ICSD) 
or the Crystallographic Open Database~\cite{COD} (COD) (the latter two, combined, contain to date over half a million entries).
In comparison, 2D materials databases are still scarce and limited in size: a
first scan of the ICSD database identified $92$ two-dimensional compounds~\cite{Nieminen}  (including CuS$_2$, subsequently synthesised~\cite{Romdhane2015}). This was 
followed by $103$ compounds
selected among specific classes~\cite{miro2014atlas}, while a recent study focused on transition-metal dichalcogenides and oxides, identifying $51$ of them as stable~\cite{rasmussen2015computational}. More extensive efforts~\cite{Hennig17,Reed17} have also been put forward to expand the set of prospective 2D materials by screening crystal structures from the Materials Project~\cite{MaterialsProject}. 
In fact, high-throughput computational methods represent a powerful tool\cite{Gould2016} to explore materials space or to screen materials  without the need to synthesise them first\cite{franceschetti1999,johannesson2002,curtarolo2003,curtarolo2013,Jain2016}. For instance, these techniques have been successfully employed in the search for novel materials for Li-air and Li-ion batteries\cite{mueller2011,OQMD}, for hydrogen storage\cite{Ozolins2008}, scintillators~\cite{Ortiz2009}, electrocatalysis\cite{Greeley2006}, or to accelerate the discovery of novel light-absorbing materials\cite{Yu2012}.

In this work,  we systematically explore experimentally known compounds for possible exfoliation, paying particular attention, at variance with some of the earlier work, to the mechanical stability of the exfoliated layers and the changes in the electronic structure that take place in reducing the dimensionality - from the emergence of magnetic order to that of charge density-wave instabilities. We perform this search starting from the inorganic compounds recorded in ICSD and COD, and then use vdW density-functional theory (DFT) simulations to test these 3D parents for 2D exfoliation. 
In particular, we compute the binding energy of 2662 prospective layered structures and identify those that are held together
by weak interactions and ready for mechanical\cite{Novoselov2004} or liquid-phase\cite{Nicolosi2013} exfoliation. This results into a portfolio of 1825 materials that can be exfoliated in monolayers or multilayers.
To showcase their potential we explore the electronic, vibrational, magnetic, and topological properties of the 258 most promising systems, disclosing a wealth of functional materials that can be studied experimentally, notably including 56 magnetically-ordered monolayers.

The reproducibility of all results is ensured by the deployment of the AiiDA~\cite{AiiDA} materials' informatics infrastructure, which keeps track of the full provenance for each calculation and result.

\section*{Results and discussion}
\subsection*{Starting set of three-dimensional structures}

The search protocol starts with the identification of a comprehensive initial set of 
bulk 3D crystal structures. These are extracted from databases of experimental compounds, and
for this study we focused on the ICSD~\cite{ICSD}  
and COD~\cite{COD} databases, containing respectively $177343$ (ICSD 2015.1) and $351589$ entries (COD Rev.~\#171462).
Entries containing at most $6$ different species are retrieved in the form of Crystallographic Information Files (CIF) using routines distributed within the AiiDA platform~\cite{AiiDA}, and filtered~\cite{CODtools} to correct typical syntax errors that prevent the successful reconstruction 
of the crystal structures (see Methods for more details). Entries with partial occupancies in the unit cell are then discarded, together with those that do not provide the explicit position of one or several atoms, those that cannot be parsed, or that are obviously incorrect. Last, theoretical structures that are present without an experimental counterpart are also removed.

These entries are then converted into AiiDA structures using the pymatgen~\cite{pymatgen}
parser, and the spglib library~\cite{spglib} is subsequently used to identify primitive cells and to refine atomic positions, in order to avoid loss of crystal symmetries due exclusively to round-off errors (see Methods for more details). 
Finally, duplicate structures are removed: first, independently, on each of the ICSD and COD sets; then, every structure of ICSD for which there exists a similar structure inside COD is also removed.
This procedure provides a starting point of $108423$ unique 3D crystal structures (set A, see Table~\ref{tab:stat}), flagged for further investigation.
\subsection*{Identification of layered compounds\label{sec:2D}}
\begin{figure*}
 \includegraphics{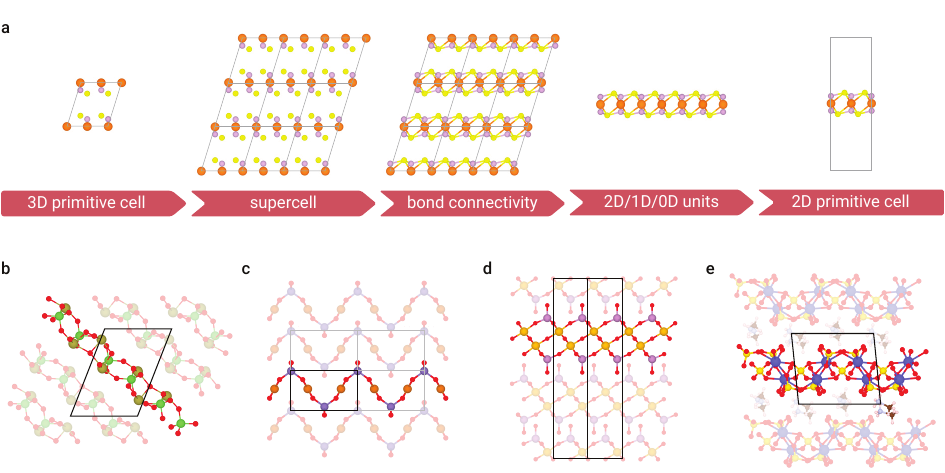}
 \caption{{\bf Screening for low-dimensional manifolds in a parent 3D crystal}: 
 {\bf a} Schematic representation of the fundamental steps needed to find low-dimensional subunits of a parent 3D crystal (here MgPS$_3$).
 {\bf b-e} Examples illustrating non-trivial layered structures that can be identified in: 
 {\bf b} Triclinic or monoclinic structures that are not layered along a standard crystallographic direction (As$_2$Te$_3$O$_{11}$, from ICSD \#425299).
 {\bf c} Layered compounds whose constitutive layers extend over multiple unit cells and that thus require the use of supercells to be identified (CuGeO$_3$, from COD \#1520902).
 {\bf d} Layers that have partial overlap of the atomic projections along the stacking direction, with no manifest vacuum separation between them (Mo$_2$Ta$_2$O$_{11}$, from ICSD \#247163).
 {\bf e} Composite structures that contain subunits with different dimensionality ( (CH$_6$N)$_2$(UO$_2$)$_2$(SO$_4$)$_3$, from ICSD \#192165, where 2D layers of uranyl sulphate are intercalated with 0D methylammonium molecules).
 \label{fig:2Dlayerfinder}}
\end{figure*}
Set A can then be  analysed to find promising candidates for exfoliation. 
The first step consists in identifying compounds that can be decomposed into bonded manifolds held together reciprocally by weak interactions. 
A pioneering attempt in this direction has been put forward by 
Leb{\`e}gue and coauthors~\cite{Nieminen} albeit focused to high-symmetry compounds having layers perpendicular to the $[001]$ crystal axis. A more versatile algorithm is nevertheless needed to cope with the following more general and realistic situations:
\begin{itemize}
\item layered structures belonging to any crystal system (obvious exceptions being the cubic systems, that cannot display a preferred exfoliation plane);
\item arbitrary orientation of the stacking direction for the individual layers (see, e.g., Fig.~\ref{fig:2Dlayerfinder}b);
\item layers' connectivity outside the 3D primitive unit cell, requiring the use of sufficiently large supercells (see, e.g., Fig.~\ref{fig:2Dlayerfinder}c);
\item layers that are interpenetrating, i.e. with overlapping projections along the stacking direction (see, e.g., Fig.~\ref{fig:2Dlayerfinder}d);
\item composite structures that include manifolds with different dimensionalities (see, e.g., Fig.~\ref{fig:2Dlayerfinder}e);
\item different chemical environments affecting the bond distances between atoms, requiring some tolerance in the parameters chosen to identify connected manifolds.
\end{itemize}

Here, we provide a protocol that is able to deal with all the scenarios above and identify possible candidates that can then be fully characterized with first-principles calculations, and verify the existence of chemically-bonded 2D monolayers held together by weak forces.

This first screening is reported schematically in Fig.~\ref{fig:2Dlayerfinder}a. First, from the bulk primitive cell of the 3D candidate a $3\times3\times3$ supercell is created. All interatomic distances are evaluated and chemical bonds are heuristically identified as those for which
\begin{equation}
 d_{i,j} < r_i^\text{vdW} + r_j^\text{vdW} - \Delta, \label{eq:bonding}
\end{equation}
where $d_{i,j}$ is the distance between two atoms $i$ and $j$, $r_i^\text{vdW}$ is
the van-der-Waals radius of atom $i$ according to the prescriptions of Alvarez (Ref.~\onlinecite{alvarez}), and
$\Delta$ is a tunable parameter. 
The criterion in~\eqref{eq:bonding} is based on the statistical analysis of the distribution of over five million interatomic distances, according to which
Alvarez pointed out that covalent bonds are likely to form only when the distance between atoms is shorter than the sum of the vdW radii by more than a quantity $\Delta$, on average equal to 1.3~\AA.
Thus, if the interatomic distance is smaller than $r_i^\text{vdW}+r_j^\text{vdW} - \Delta$ the 
two atoms should be considered chemically connected. On the contrary, if 
the interatomic distance is larger, then strong chemical bonding is absent	
and only vdW interactions remain. Some special case involving hydrogen or metallic bonding are still possible and discussed in
Ref.~\onlinecite{alvarez}, but we will not consider them here; we have considered instead five equally spaced values for $\Delta$ between $1.1$~\AA\ and $1.5$~\AA\, to account for possible uncertainties in the values of the vdW radii and of $\Delta$ itself, and to maximize the number of candidates that would undergo the ultimate electronic-structure screening.
We believe that criterion to be more robust and versatile than one based on covalent radii, as it allows to take into account the bond-length fluctuations associated with different chemical environments. 

Once all bonds are identified, chemically connected groups are constructed. We start from $3\times3\times3$ supercells, as illustrated in Fig.~\ref{fig:2Dlayerfinder}c (even such supercells might not be large enough in principle), and identify
the dimensionality of each connected manifold  from the rank of the matrix formed by all the vectors linking an atom to all of its chemically connected periodic images (periodic according to the 3D lattice vectors).
This approach is general and does not assume any specific orientation of the 2D plane with respect to the crystal or Cartesian axes. Moreover, it also identifies and returns other low-dimensional units, such as 1D chains or 0D clusters. 
Applying the full protocol to all $108423$ 3D structures of set A, we find $5619$ materials that are classified as layered according to this simple geometric and chemical analysis (set B in Table~\ref{tab:stat}) that can be considered for further analysis.

\subsection*{Structural relaxation and comparison with
experiments\label{sec:relax}}
\begin{figure*}[ht]
 \includegraphics{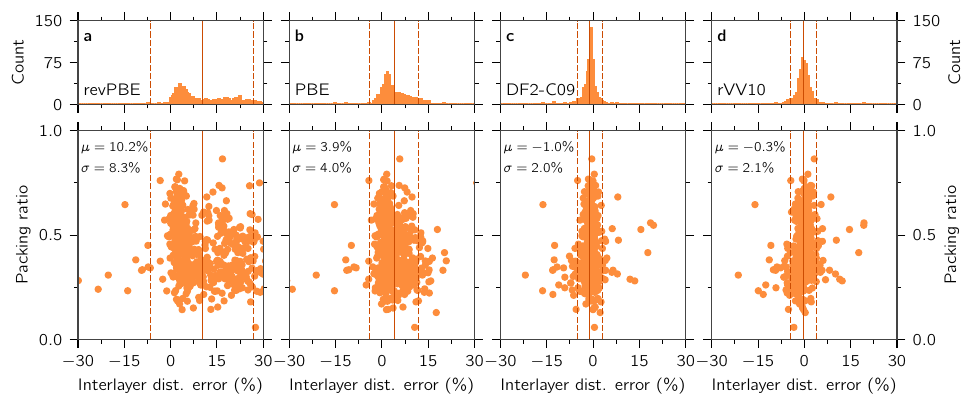}
 \caption{{\bf DFT structural relaxations vs experiments}:
Top panels: distribution of the relative error in the interlayer distance calculated with different energy functionals with respect to the value for the corresponding experimental structure, for 573 layered materials
 in set B for which the structure has been explicitly marked as experimental in the parent database, and for which a relaxation has been performed with all four exchange-correlation functionals.
The four functionals considered are:
 {\bf a} the revPBE functional,
 {\bf b} the PBE functional,
 {\bf c} the DF2 non-local vdW functional with C09 exchange,
 {\bf d} the revised VV10 non-local vdW functional. 
Vertical solid lines are located at the distribution average $\mu$, while dashed lines are located two standard deviations $\sigma$ away from the average (we removed outliers when calculating the position of these lines, outliers being defined as data points sitting in bins with less than three counts, with a bin width of 0.6\%).  Lower panels:
Scatter plot of the packing ratio versus the interlayer distance error. The values of the average $\mu$ and standard deviation $\sigma$ are also reported.
\label{fig:volume_vs_exp}}
\end{figure*}
This geometrical characterization of set B is the starting point for accurate electronic structure calculations to identify materials that exhibit weak interlayer bonding and could be exfoliable. Here we use vdW DFT simulations to proceed with such a refinement (computational details can be found in the Methods).

The first step is the optimization of the 3D geometry
(cell vectors and atomic positions) of the structures found. 
This procedure is completely automated within AiiDA.
Preference is given to the simplest structures, both for their higher relevance
and for computational efficiency. As a result, at least all unary compounds containing less than $100$~atoms in the unit cell,
all binaries and ternaries with less than $40$~atoms, 
and all other quaternary, quinary, and senary compounds with less than $32$~atoms are considered; in total, $3210$ layered compounds (set C in Table~\ref{tab:stat}). These structures are relaxed using two different non-local vdW functionals:  the vdW-DF2 ~\cite{lee2010} with C09 exchange~\cite{cooper2010,hamada2010} (DF2-C09) and the revised Vydrov-Van Voorhis~\cite{vydrov2009,vydrov2010,sabatini2013} (rVV10) functional.

To assess the accuracy in predicting the equilibrium structure for the functionals adopted, we compare the experimental interlayer distance (defined as the distance between the geometrical centres of neighbouring layers) of a subset of layered materials against computational predictions.
The results are shown in Fig.~\ref{fig:volume_vs_exp}, complemented also by the semi-local Perdew-Burke-Ernzerhof (PBE)\cite{PBE} and the revPBE\cite{revPBE} functionals.
Since  revPBE and PBE are not designed to predict vdW interactions, discrepancies with experiments are larger, as seen in Fig.~\ref{fig:volume_vs_exp}a-b.
On the contrary, the two vdW-compliant functionals provide interlayer distances  closer to experimental values and with a much smaller dispersion around the average (see Fig.~\ref{fig:volume_vs_exp}c-d), and are able to correctly predict the
structural properties of non-closely-packed materials (including vdW-bonded structures).
This can be quantified by looking at the mean absolute percentage errors ($1.5\%$ for both vdW-DF2-C09 and rVV10, against $10\%$ and $4\%$ for revPBE and PBE, respectively).
Moreover, the  low mean relative errors of vdW-DF2-C09 and rVV10 predictions ($-1\%$ and $-0.3\%$, respectively) show that these functional have no bias in determining the equilibrium interlayer distance of layered structures, while revPBE and PBE have a tendency to underbind that is emphasised at smaller packing ratios.

\subsection*{Binding energies  and final selection of exfoliable materials\label{sec:binding}}
\begin{figure*}
\includegraphics{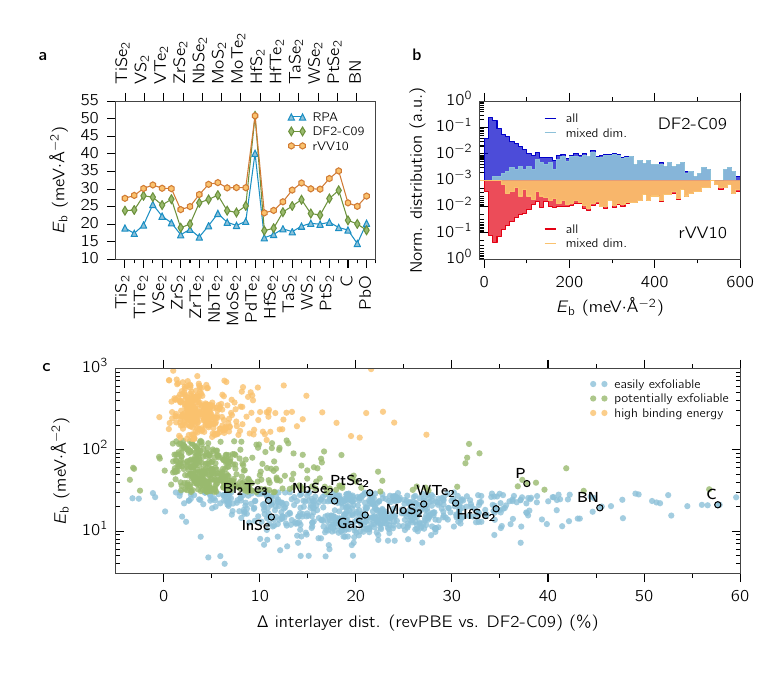}
\caption{{\bf Binding energies} of the bulk 3D compounds identified as geometrically layered:
 {\bf a} Binding energy for a selection of layered materials identified in Ref.~\onlinecite{bjorkman2012},
comparing results calculated with the random-phase approximation (RPA) (Ref.~\onlinecite{bjorkman2012}) and from the DF2-C09 and rVV10 functionals (this work).
 {\bf b} Normalised distribution of the all the binding energies computed in this work (set D in Table~\ref{tab:stat}).
 {\bf c} Binding energy vs change in interlayer distance between the structure relaxed using the revPBE and the DF2-C09 functionals.
Materials classified as easily exfoliable, potentially exfoliable, or with high binding energies are reported in different colours. Well-known 2D materials are highlighted.
\label{fig:binding}}
\end{figure*}
To go further in the selection of promising materials, we
need to determine if the interlayer interactions are weak enough for the
parent compound to be a good candidate for exfoliation.  For this, we calculate the overall binding energy $E_\text{b}$ as the difference in 
the ground-state energy  between the 3D bulk and all of its substructures (of any dimensionality, including also 1D chains or 0D clusters)~\cite{bjorkman2012}. In order to validate the quality of the vdW-DF2 and rVV10 functionals, we show in Fig.~\ref{fig:binding}a the 
binding energies per layer and per unit surface calculated for well-known layered materials including 
graphite, hexagonal boron nitride and several transition-metal dichalcogenides. It can be seen that the values
obtained with the two vdW functionals are in very good agreement
with accurate reference calculations performed in Ref.~\onlinecite{bjorkman2012} using the random-phase approximation (RPA).
 
We then show in Fig.~\ref{fig:binding}b the distribution of binding energies for the $2662$  structures that are still flagged as layered after vdW-DFT relaxation of set C; we label these as set D in Table~\ref{tab:stat}. 
For both vdW functionals, a majority of compounds ($74\%$ for DF2-C09 and $71\%$ for rVV10)
exhibit a binding energy smaller than $100$~meV$\cdot\text{\AA}^{-2}$, giving rise to the first  peak in the distribution (note the logarithmic scale). 
A second broader peak appears
between $200$ and $400$~meV$\cdot\text{\AA}^{-2}$ and is associated with 
structures containing also lower dimensionality units (i.e., 0D or 1D) in addition to the 2D layer(s).
The larger binding energy is to be expected for structures with mixed dimensionality (typically 0D and 2D) where 
charge transfer between intercalated units and 2D layers can take place giving rise to ionic bonds.
This means that while these additional 0D or 1D substructures   have been separated  from the 2D layers in the geometrical screening, the parent compounds are actually not exfoliable. 

The binding energies distribution is quite continuous. As a result, a criterion to select layered materials solely based on binding energies 
is difficult to define without a level of arbitrariness. To lift this ambiguity, we correlate the binding energy with
the relative discrepancy in equilibrium interlayer distance obtained in the structural relaxation performed with vdW and non-vdW functionals, using for the latter revPBE~\cite{revPBE} that, contrary to standard PBE, does not seem to display bonding interactions in the exchange energy~\cite{Dion2004}. 
We plot in Fig.~\ref{fig:binding}c  the binding energies against this interlayer-distance difference for $1535$ compounds randomly chosen from set D that have been studied  with the revPBE functional in addition to the DF2-C09 functional, and in Supplementary Figure 1 for the 1482 compounds  that have been studied with revPBE in addition to rVV10. In both cases
we observe that most compounds with low binding energies exhibit a large relative discrepancy in interlayer distance calculated with the two functionals. This is expected, since the interlayer interaction is very sensitive to vdW forces but it also allows us to identify much more clearly which layered compounds are characterised by dispersive interactions between layers, and thus to set robust thresholds of {$30$~meV$\cdot\text{\AA}^{-2}$ for binding energies computed with the DF2-C09 functional, and $35$~meV$\cdot\text{\AA}^{-2}$ for those computed using rVV10. We classify  parent compounds falling below one or both of these thresholds  as ``easily exfoliable'' (EE), shown in blue in Fig.~\ref{fig:binding}c and in Supplementary Fig.~1. This name is also applied to the resulting monolayers. As shown in Fig.~\ref{fig:binding}c  and in Supplementary Fig.~1, this choice recovers 2D materials commonly exfoliated in experiments~\cite{Nicolosi2013}, validating the approach.

At the top left of Fig.~\ref{fig:binding}c and in Supplementary Fig.~1, a number of compounds exhibit high binding energies and a similar interlayer distance when comparing the revPBE and vdW functionals; many also contain substructures of mixed dimensionalities, as discussed before. This group can be clearly separated from the rest, and is shown in yellow, the boundary being set at $130$~meV$\cdot\text{\AA}^{-2}$ for both vdW functionals. Above this value, compounds are  not considered exfoliable and they are discarded from the database.

Between these two regions, the remaining compounds have intermediate binding energies (shown in green) and exhibit relatively weak, possibly non-vdW, bonding. As an example, the compound PdTe$_2$ belongs to this region, and is metallic out-of-plane. We classify parents belonging to this group and the resulting monolayers as ``potentially exfoliable'' (PE).
Finally, materials for which the binding energy has been computed with both vdW functionals are classified according to the most optimistic prediction.

We note in passing that 112 parent compounds can be exfoliated into more than one different monolayers.

\begin{table*}[t]
\begin{tabular}{lcccr}
\multirow{2}{*}{} & Unique & Unique & Common & Total \\
 & to ICSD & to COD & to both & sum \\
\hline\hline
Experimental data &&&&\\\hline
CIF inputs & 99212 & 87070 & & 186282 \\\hline
Unique 3D structures (set A)& 34548 & 60354 &13521 & 108423 \\ \hline
Layered 3D structures (set B) & 3257 & 1180 & 1182 & 5619 \\\hline\\\hline\hline
\multicolumn{5}{l}{DFT calculations } \\\hline
Layered 3D, relaxed (set C) & 2165 & 175 & 870 & 3210 \\\hline
Binding energies (set D) & 1795 & 126 & 741 & 2662 \\\hline
2D easily exfoliable (EE) & 663  & 79 & 294 & 1036 \\
2D pot. exfoliable (PE) & 524 & 34 & 231 & 789 \\
Total & 1187 & 113 & 525 & 1825 \\ \hline  
\end{tabular}
\caption{{\bf Database statistics}.  Experimental data: number of structures imported from the two databases (ICSD or COD, uniqueness not tested), number of unique 3D structures in each imported set or common to both (set A), and number of layered 3D structures identified using the geometrical criteria discussed in the text (set B). DFT calculations: number of structures that were relaxed (set C), number of structures that  remain classified as layered after relaxation and for which binding energies were computed (set D), and number of easily or potentially (see text) exfoliable compounds.\label{tab:stat}}
\end{table*}

\subsection*{The 2D database}
\begin{figure*}[t]
 \includegraphics{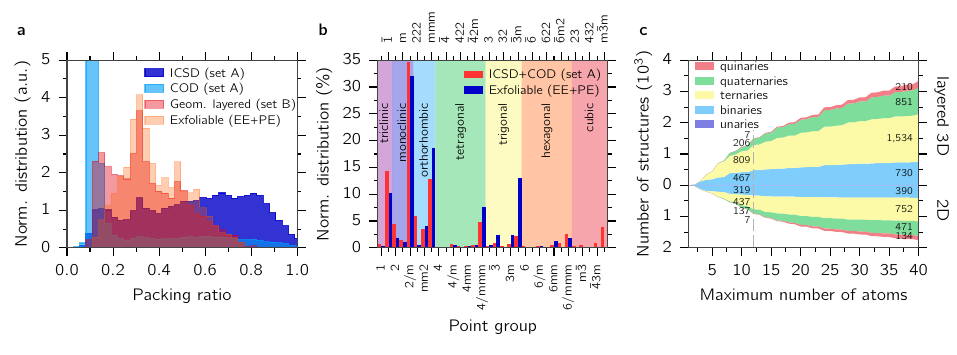}
 \caption{{\bf: Statistics on the 2D and 3D databases}:
 {\bf a} Packing ratios for the 3D structures considered according to their ICSD or COD provenance, layered compounds (set B) and exfoliable materials (EE+PE).
 {\bf b} Distribution of point groups.
 {\bf c} Number of structures as a function of the number of atoms in the primitive cell (top: layered 3D structures; bottom: 2D structures in the EE and PE groups combined). The number of structures with at least two different species is explicitly reported for a maximum number of atoms in the primitive cell equal to 12 and 40. The number of unary structures with less than 12 (40) atoms in the primitive cell is 13 (15) for 3D layered compounds and 15 (15) for 2D EE+PE materials.
 \label{fig:2Dstats}}
\end{figure*}
Using these criteria, and after removal of duplicates (that is, identical 2D materials that come from different parents, see Methods), we obtain $1036$ easily exfoliable and $789$ potentially exfoliable compounds, for a total of $1825$ candidates.
In Fig.~\ref{fig:2Dstats} we plot some statistics for this portfolio of promising materials, as well as for their 3D parents. We first show the distribution of packing ratios in the ICSD and COD structures (set A), in the group of 3D layered compounds (set B), and in the EE and PE ones, using  covalent 
radii~\cite{cordero} to compute the atomic volume. While the initial databases contain many closely packed
structures with packing ratios of 0.7 or more, the layered structures are peaked around  low packing ratios (0.3) but with a spread extending up to 0.8, even in the final set of exfoliable compounds. These considerations highlight how the packing ratio alone is not necessarily a good criterion to filter 2D structures (note that the peak of COD at a packing ratio of 0.15 -- which extends beyond the upper limit of the graph -- is due to the fact that COD is populated with a large number of molecular crystals, microporous networks, metal-organic frameworks, and zeolites).

Then, in Fig.~\ref{fig:2Dstats}b we show the distribution of point groups,  arranged by crystal system, in all 3D structures, compared to that of the exfoliable 3D compounds. Most point groups are represented similarly in both sets, with the notable exception of all  cubic point groups that are obviously absent from the set of layered compounds.
Moreover, the $222$ point group  is much rarer in the layered compounds than in the ICSD and COD databases, while the $3$m, $\bar{3}$, $\bar{3}$m and $6$mm point groups are conversely  more present among 3D exfoliable compounds.
It is interesting to note that while most of the 2D materials studied in the last ten years belong to the hexagonal crystal system, these structures represent only a minority of the full database of exfoliable materials.

Finally, in Fig.~\ref{fig:2Dstats}c we show how the layered and 2D compounds are distributed in terms of the number of species and number of atoms in the unit cell. 
As apparent from the graph, there are hundreds of 2D structures with few atoms in the unit cell ($258$ EE and $116$ PE with at most $6$ atoms, $591$ EE and $324$ PE with at most $12$ atoms), that can be used as a starting set for screening 2D materials with desired properties.
We find 18 unary monolayers (i.e. containing a single species), including all known exfoliable 2D unaries such as graphene and phosphorene. Novel unaries belong to the class of PE materials and mostly arise from intercalation layers in the parent 3D structures, although they are typically not mechanically stable.
Interestingly, ternary 2D materials, which have been so far largely unexplored, represent a significant fraction of structures with less than 12 atoms in the unit cell, demonstrating that it may be worth inspecting these more versatile compounds beyond the current realm of unary/binary 2D materials.

\begin{figure*}[t]
 \includegraphics{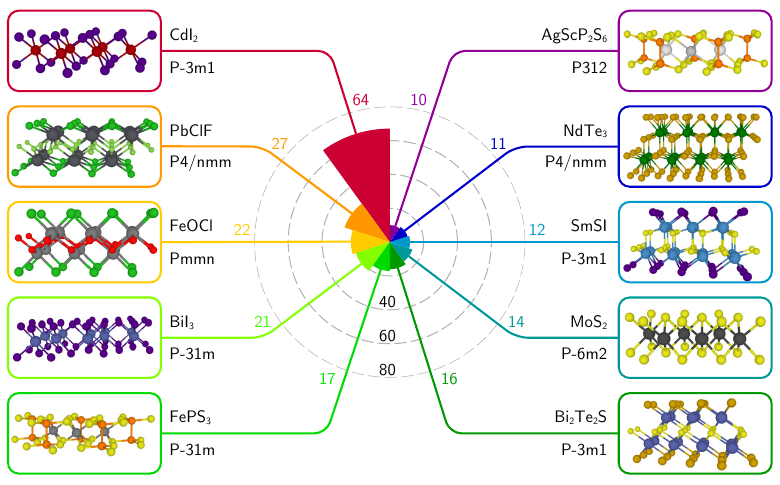}
 \caption{{\bf Most common 2D structural prototypes}: Polar histogram with the number of structures belonging to the ten most common 2D structural prototypes in the set of 1036 easily exfoliable 2D materials. A graphical representation of each prototype is reported, together with the structure-type formula and the spacegroup of the 2D systems.\label{fig:2Dprototypes}}
\end{figure*}

In order to provide a more precise overview, we further classify the 2D materials of the EE group into different prototypes, according to their spacegroups and their structural
similarity, when species are all considered undistinguishable and only the
stoichiometry matters (for this we use the pymatgen structure matcher -- see Methods for more details). $562$ prototypes are found, among which the ten most common ones represent a total of $214$ structures and are shown in Fig.~\ref{fig:2Dprototypes}. The most common structural prototype is that of CdI$_2$, which groups together $64$ similar structures, including many transition-metal dichalcogenides and dihalides. Contrary to the general distribution of point groups in Fig.~\ref{fig:2Dstats}b, most of these prototypes have high-order rotation axes, with the only exception being the rectangular FeOCl. Although some of these structures have already been suggested as exfoliable~\cite{Nieminen}, the abundance of some relatively complex prototypes involving more than two species is  new and compelling. Moreover, we emphasise that the NdTe$_3$ prototype has not been reported before and it is common to many rare-earth tritellurides.

 Buoyed by these results, we extensively characterize here the most promising 2D structures, i.e. the 258 easily exfoliable compounds with at most 6 atoms per unit cell.  For this set we study  the mechanical stability and vibrational properties together with their electronic structure, with particular care on all possible possible ferromagnetic or antiferromagnetic states, in order to verify the mechanical and electronic ground state of the exfoliated monolayers.  Remarkably, we find a number of very promising materials, including 37 ferromagnets (of which 13 insulators), 19 antiferromagnets (of which 12 insulators), 6 systems exhibiting sharp unstable phonon modes that could correspond to charge density waves, 2 quantum spin Hall insulators, and 4 compensated semimetals; all of these are presented in the Supplementary Material. Such results hint at the usefulness of high-throughput studies in the search for novel high-performance 2D materials.

As mentioned, we start by assessing structural stability, as the existence of 2D subunits in a 3D parent compound does not guarantee that the isolated monolayer would still be stable. We therefore compute the phonon frequencies at $\Gamma$ and, if we find imaginary-frequency modes, we displace atoms (see details on the procedure in the Methods section) along the modes eigenvectors and then fully relax the structure (this approach is used, rather than force relaxations, to avoid saddle points due to symmetries).  The procedure is iterated until all phonon energies at $\Gamma$ become real and positive. We then compute the full phonon dispersions in the correct 2D (out-of-plane open-boundary conditions) framework (see Methods). This is extremely relevant and necessary to correctly describe optical phonons in polar materials~\cite{sohier2016a}. All computed phonon dispersions are displayed in the Supplementary Material (for magnetic materials, the phonon dispersions are computed on the spin-polarized ground state configuration -- see below). We emphasize that, by limiting the stabilization procedure described above to imaginary phonons at $\Gamma$ only, we do not take into account mechanical instabilities associated with supercell structural displacements. Nevertheless, these can be detected by inspecting the full vibrational spectrum, that encompasses important clues to disclose emerging phenomena, including soft modes or charge-density-wave instabilities, like in the case of CoI$_2$. For these cases, full stabilization could be obtained by using the same procedure described above, considering also modes associated to imaginary phonon frequencies outside $\Gamma$ (requiring to work in a supercell).

More than 80\% of the systems studied exhibit positive phonon dispersions,  supporting the soundness of the exfoliation protocol. To further characterize these 2D materials, we also report their electronic band structure along high-symmetry lines. Most of the structures (166) are semiconducting, with 78 having band gaps ranging from 0 to 1.5~eV (at the PBE level), while 92 are metallic.

Recently, interest in magnetic 2D materials has exploded~\cite{Samarth2017} owing to the isolation of the first ferromagnetic mono- and bilayers~\cite{Huang2017,Gong2017}. To expand this family, we fully characterize the electronic ground state by comparing the energy of several magnetic configurations (including both ferromagnetic and antiferromagnetic supercell ordering) and hence identify the magnetic ground state of each 2D material. Mechanical stability is assessed on these magnetic structures by computing the phonon dispersions in the most stable magnetic configuration found, and when the phonons at $\Gamma$ are found negative, the structure is stabilized in the same way as described above. A total of 56 compounds are found to support non-trivial magnetic ordering and are reported in Table~\ref{tab:magnetism}, including magnetic insulators, half-semiconductors (e.g. CoCl$_2$ and CrBrO), and half-metals (e.g. FeBr$_2$), with great promises for spintronic and data-storage applications. The lowest-energy magnetic configuration and its corresponding magnetization for these 258 monolayers are reported in Supplementary Tables 1--3. We note that the recently synthesised ferromagnetic semiconductors CrI$_3$~\cite{Huang2017} and CrGeTe$_3$~\cite{Gong2017} are not present in this list as they contain more than 6 atoms per cell, although the database contains many more prospective magnetic 2D materials with the same structural prototype (BiI$_3$ and FePS$_3$ in Fig.~\ref{fig:2Dprototypes}). 

Quantum spin Hall insulators (QSHI) display $\mathbb{Z}_2$ topological order and have often been proposed as a platform to realize novel dissipations-less devices for electronics and spintronics. By performing DFT-PBE calculations and following the evolution of the hybrid-Wannier-function centers (see Methods), we identify only two QSHI in this subset of EE compounds. The first QSHI is made of pure bismuth with space group 164 (P-3m1) and 2 atoms/unit cell, also known as bismuth-bilayer or bismuthene \cite{BismuthBilayer}. The second one is TiNI \cite{TiNI-physchem}, a ternary compound with space group 59 (Pmmn) and 6 atoms/unit cell, exhibiting a band inversion between iodine p-orbitals and titanium d-orbitals gapped by spin-orbit coupling. In addition, we found systems (such as WTe$_2$ or MoTe$_2$) that are metallic at the DFT-PBE level, but whose band structure could in principle be adiabatically connected to a QSHI, e.g via a small strain or the interaction with a substrate.
\begin{table}
\setlength{\tabcolsep}{5pt}
\begin{tabular}{ r p{0.3\linewidth} p{0.3\linewidth}}
 & \centering Ferromagnetic & \centering Antiferromagnetic \tabularnewline
\hline
Metals & CoH$_{2}$O$_{2}$, CoO$_{2}$, ErHCl, ErSeI, EuBrO, EuIO, FeBr$_{2}$, FeI$_{2}$, FeTe, LaCl, NdBrO, PrBrO, ScCl, SmBrO, SmSI, TbBr, TmI$_{2}$, TmIO, VS$_{2}$, VSe$_{2}$, VTe$_{2}$, YCl, YbBrO, YbClO & CoI$_{2}$, CrSe$_{2}$, FeClO, FeO$_{2}$, FeSe, PrIO, VBrO \\
\hline
Semiconductors & CdClO, CoBr$_{2}$, CoCl$_{2}$, CrBrO, CrClO, CrSBr, CuCl$_{2}$, ErSCl, HoSI, LaBr$_{2}$, NiBr$_{2}$, NiCl$_{2}$, NiI$_{2}$ & CrBr$_{2}$, CrI$_{2}$, LaBr, MnBr$_{2}$, MnCl$_{2}$, MnH$_{2}$O$_{2}$, MnI$_{2}$, VBr$_{2}$, VBr$_{2}$O, VCl$_{2}$, VCl$_{2}$O, VI$_{2}$ \\
\hline
\end{tabular}
\caption{{\bf Easily-exfoliable magnetic compounds}: Chemical formulas of the easily-exfoliable magnetic monolayers with at most six atom in the unit cell.
Compounds are grouped according to the type of magnetic ordering and the presence of an energy band gap in the electronic structure.\label{tab:magnetism}}
\end{table}

\section*{Conclusions\label{sec:concl}}
We identified an extensive database of exfoliable materials by systematic analysis and screening of experimental structures extracted from both the ICSD and the COD structural databases. 
We developed a geometrical algorithm to select potentially layered materials that is general enough to recognise substructures independently of their crystallographic orientation, dimensionality and environment. 
The list of 2D materials obtained by geometrical considerations is then tested against accurate, validated  vdW-DFT computations
of the binding energies for all layers identified. 
Compounds with sufficiently small binding energies and where dispersion relations affect interlayer distances are deemed exfoliable. The materials identified  are in particular classified into groups of easily ($1036$) or potentially ($789$) exfoliable compounds, showing that only a very small fraction of possible 2D materials has been considered up to now.
In particular, many opportunities might arise from materials with reduced symmetry or involving more than two atomic species but still with relatively simple structures;
the identification of the most common prototypes could also allow to further expand the list of novel 2D materials by chemical substitutions and alternative site decorations.

For easily exfoliable compounds with up to 6 atoms/cell, we comprehensively characterize their stability, vibrational, electronic, magnetic, and topological properties, revealing a wealth of magnetic systems (56 magnetically ordered, including 37 ferromagnets, 19 antiferromagnets, 14 half metals, and 6 half semiconductors) and highlighting a relative scarcity of insulators with $\mathbb{Z}_2$ topological order.
All results can be fully reproduced thanks to the deployment of the AiiDA infrastructure that tracks the provenance of each data entry.
To date, this is the largest available database of 2D compounds and will soon be made available online on the Materials Cloud platform (at http://www.materialscloud.org).

\section*{Methods\label{sec:methods}}

\subsection*{Reproducibility and provenance\label{sec:provenance}}

It is an often repeated tenet that results of scientific research must be reproducible. This objective is, however, challenging, especially in high-throughput research, due to the large number of simulations involved and the complex sequence of logical steps needed in the study. 
To ensure reproducibility we use AiiDA~\cite{AiiDA} as a materials' informatics infrastructure to implement the ADES model of Automation, Data, Environment, and Sharing, as discussed in Ref.~\onlinecite{AiiDA}. All input information for each calculation is stored in the AiiDA repository before calculations are launched and actual code inputs are created by AiiDA using only data already stored in the repository. 
However, even the ability to replicate single calculations is not sufficient, since the  majority of results are obtained via long sequences of simulations. It is thus essential to persist  the full provenance of each data item. This means, for instance, that for any 2D material identified in this work one can track from which 3D parent structure it originated (including its source database), how the structure was processed to obtain its  coordinates from the Wyckoff positions, how the low-dimensional units were identified, and which parameters were used to obtain the DFT results. 

\subsection*{CIF files reformatting\label{sec:ciffilter}}

The CIF files~\cite{CIF} provided in crystallographic databases sometimes contain syntax errors or unnecessary information that needs to be corrected and/or removed before parsing them. To do so, we use cod-tools\cite{CODtools}, in particular the \texttt{cif\_filter} and \texttt{cif\_select} scripts, in order to:
\begin{itemize}
\item fix common CIF syntax errors,
\item parse and reformat the summary chemical formula (from the tag \texttt{\_chemical\_formula\_sum}),
\item set explicit $90$ degrees cell angles if they are not given (following the standard CIF specifications),
\item remove empty non-loop tags,
\item remove CIF data blocks without coordinates,
\item convert non-standard CIF tag name capitalization into the capitalization specified by the CIF dictionary, for
example \texttt{\_atom\_site\_cartn\_x} would be replaced by \texttt{\_atom\_site\_Cartn\_x}.
\end{itemize}

We also remove the tags \texttt{\_publ\_author\_name} and \texttt{\_citation\_journal\_abbrev}, that often contain problematic characters for, e.g., the pymatgen CIF parser (this information is still stored in the AiiDA database and therefore fully searchable).

As already mentioned in the main text, in this step we consider only structures coming from experimental measurements; nevertheless the source databases are partially populated by purely theoretical structures. A selection is made using the flags set on the database entries by their curators.
We also implemented some heuristics to detect clearly incorrect CIF files. For example, we discard structures where the chemical formula provided in the file is inconsistent with the elements in the unit cell. Regardless of all these efforts, it is possible that some incomplete or incorrect structures are still not filtered out from the original databases.

\subsection*{Crystal structure refinement\label{sec:refine}}

The crystal structures parsed from CIF files are often subject to round-off errors
in the atomic positions and/or cell parameters that artificially lower the
number of symmetries. We use the spglib software~\cite{spglib} to refine 
each structure and recover the maximum number of symmetries,
using the algorithm described in Ref.~\onlinecite{Grosse-Kunstleve}. Interatomic
distances and cell parameters are modified by at most $5\cdot 10^{-3}$~\AA{} with
respect to the initially parsed structure. At the same time (and with the same software),
a standardised crystallographic primitive cell is extracted
\cite{spglib_std}.

\subsection*{Removal of duplicates and prototyping\label{sec:duplicate}}

For the duplicate filtering procedure of both 3D and 2D structures, as well as for the prototyping of 2D structures, we have used extensively the ``structure matcher'' of pymatgen~\cite{pymatgen}, which, thanks to the algorithm described in Ref.~\onlinecite{CMPZ}, compares crystal structures. The goal is to 1) reduce the two structures to be compared to their primitive cells, 2) rescale the volumes (if 3D), or the in-plane areas (if 2D), 3) find, when it exists, an affine map between the two cells, within certain tolerances on the cell lengths and angles, and 4) compute the maximum  distance between atomic sites when the two lattices are mapped onto a common average lattice. If this maximum is lower than a certain fraction of the average free length per atom, then the two structures are classified as similar.

When comparing 3D structures, the  tolerances on cell lengths, cell angles, and distances between paired atomic sites were chosen to be   $20\%$,  $5^\circ$ and $30\%$ of the average free length per atom, defined as the cubic root of the volume per atom. For 2D structures, due the  artificial vacuum added to the cell in the out-of-plane direction the relative tolerances needed to be set tighter, and we chose $10\%$,  $1^\circ$ and $10\%$, respectively.

To filter out duplicates of a set of structures, all compounds with same composition (i.e., the same species in the same relative proportion) are compared among them using the structure-matcher algorithm, forming groups of matching structures. One unique structure is then chosen in each group as a representative for further calculations.

The same tool is used to prototype 2D structures, but with a number of modifications: 1) two structures must have the same space-group to be considered as members of the same prototypical class (we used spglib with a precision parameter of $0.5$~\AA\ to get the space-group, see above), 2) the comparison is made irrespective of the specific species on each atomic site (e.g., MoS$_2$ and WTe$_2$ are both considered as AB$_2$), 3) when the two primitive cells do not contain the same number of sites, one also attempts to generate a supercell of the smaller lattice before the mapping, and 4) the tolerances on respectively the cell lengths, cell angles and maximum distance between sites are set to $10\%$,  $5^\circ$ and $20\%$ respectively. Again all structures of the same kind of composition (e.g., AB$_2$, A$_2$B$_3$, etc.) are compared in pairs, to classify them into distinct prototypical classes.

\subsection*{DFT calculations\label{sec:DFT}}

We use the Quantum ESPRESSO distribution\cite{QE-2009} with the SSSP efficiency pseudopotentials' library (version $0.7$)~\cite{SSSP}. This library of extensively tested pseudopotentials from various sources\cite{GBRV,pslib031,pslib100,SG15,Goedecker,Wentzcovitch} provides to date the best overall agreement with respect to all-electron calculations\cite{Cottenier2016,Lejaeghere2014}. Wavefunction and charge-density cutoffs are chosen according to convergence tests with respect to  cohesive energies, stresses and phonon frequencies performed for each individual element\cite{SSSP}.

Two different van-der-Waals functionals are employed, namely the vdW-DF2-C09\cite{lee2010,cooper2010,hamada2010} and the revised VV10\cite{vydrov2009,vydrov2010,sabatini2013} that have been proved to perform very well in layered systems\cite{hamada2010,bjorkman2012,Bjorkman2012prb,Bjorkman2014}, in addition to the PBE\cite{PBE} and revPBE\cite{revPBE} functionals. These vdW functionals have a complex functional form~\cite{lee2010,vydrov2010,Berland2015} with a non-local dependence on density and parameters determined to reproduce specific functional dependencies or very accurate theoretical results. Sampling of the Brillouin zone is set using a $\Gamma-$centred Monkhorst-Pack
grid, with an even number of $k$ points in each direction of the reciprocal lattice
 such that the spacing between two consecutive $k$ points along each direction is
as close as possible to $0.2$~\AA$^{-1}$. All relaxations and binding energies of 3D structures are computed using a Marzari-Vanderbilt cold smearing\cite{cold} of 0.02~Ry, and without considering spin polarization nor spin-orbit coupling (note that the binding energy of magnetic structures is not significantly affected by spin-polarization -- see Supplementary Figure 2). When computing the ground-state energy of isolated 2D units, a vacuum space of 20~\AA{} is set along the orthogonal direction to remove any fictitious 
interaction between periodic images of the 2D layers, while the atomic positions and in-plane cell parameters are kept fixed at the bulk values.
In this particular case, because the energy of the 2D layer is directly compared with the energy of the parent bulk, 
it is important to use the same computational framework in 3D and 2D calculations.

All other electronic and vibrational properties of the 258 2D materials in the portfolio are computed in a different setup. Namely, the following aspects are changed with respect to the previous setup: 1) the functional, now set to PBE, 2) the smearing, used only for metals while either totally suppressed or decreased by two orders of magnitudes for insulators, 3) the use of spin-polarization for all the systems identified as magnetic (see below), and 4) the use of the newly developed \cite{Sohier2017} 2D Coulomb cutoff for DFT and density-functional perturbation theory within Quantum ESPRESSO.  This latter approach suppresses all interactions between periodic images, leading to the correct bidimensional framework without the need for large supercell  dimensions in the non-periodic direction, which reduces the computational cost. Most importantly, the Coulomb cutoff approach is essential as soon as responses to long-wavelength perturbations are involved \cite{Sohier2016,sohier2016a}, such is the case for optical phonons in polar materials (and thus also acoustic phonons in piezoelectric materials).

The structural stabilization procedure follows the algorithm described in Ref.~\onlinecite{Togo2013}. In particular, phonons at $\Gamma$ are computed (using the 2D Coulomb cutoff described above); one group of degenerate imaginary-frequency phonons is selected (the one with largest magnitude of the imaginary frequency), and a new cell is built where atoms are displaced following the eigenvectors of the unstable mode, choosing a maximum atomic displacement of 0.11~\AA. If the mode is degenerate, a search is done on all possible linear combinations of the two modes; the displacements with largest number of symmetries are chosen. The cell is standardized using spglib\cite{spglib,spglib_std} (with a symmetry precision of 0.05~\AA), rotated if needed to have the non-periodic direction along $z$, and finally fully relaxed. The whole algorithm is repeated until all phonons at $\Gamma$ are zero or positive real (this is needed when more than one group of degenerate unstable phonons is present).

The magnetic ground state of 2D compounds is identified thanks to an automated workflow which proceeds at first by screening purely ferromagnetic states, setting random initial spin states on each atomic species (identical for all the sites occupied by the same species). If after relaxation with spin-polarized DFT, the compound ends up in a non trivial magnetic state, this step is followed by a more complete screening of all possible antiferromagnetic states, performing calculations on supercells of twice the size of the unit cell when needed, while avoiding symmetry redundant configurations (we use the algorithm of Refs.\cite{Hart1,Hart2,Hart3}, through its interface with pymatgen\cite{pymatgen}). Among all the configurations tested, the one with minimal energy per formula unit is selected.

QSHI are identified by first-principles calculations of the $\mathbb{Z}_2$ invariant using Z2pack\cite{Z2pack2011,Z2pack2017}, Wannier90\cite{Wannier90code2014} and Quantum ESPRESSO \cite{QE-2009}. In the latter case calculations are performed with the PBE functional and fully-relativistic optimized norm-conserving Vanderbilt (ONCV) pseudopotentials~\cite{oncv2013} from the PseudoDojo library~\cite{PseudoDojo,Cottenier2016}. The screening for QSHI is not performed on compounds containing lanthanides, due to the limited accuracy of standard DFT band structures in such cases.

\section*{Acknowledgements}
This work was supported by the MARVEL National Centre of Competence in Research of the Swiss National Science Foundation. Simulation time was provided by the Swiss National Supercomputing Centre (CSCS) under project IDs s580, mr0, and ch3, amounting to 60,000 calculations and $5$ million core hours. The authors would also like to acknowledge useful discussions with Francesco Ambrosio, and thank Matteo Giantomassi, Michiel J. van Setten and Gian Marco Rignanese for providing fully-relativistic ONCV pseudopotentials.
\section*{Author contributions}
M.G., G.P., N.Mo., and N.Ma.\ conceived the project.
N.Mo., A.C., G.P., A.Me., T.S., and I.E.C.\ provided necessary inputs, software tools, and AiiDA workflows.
N.Mo., P.S., and M.G.\ extracted and refined the structures from the source databases.
P.S., N.Mo., and M.G.\ performed the geometrical screening of layered materials. 
N.Mo., D.C., and A.Ma.\ performed all first-principles simulations.
N.Mo., M.G., G.P., D.C., A.Ma., and N.Ma.\ analysed the data.
All authors contributed to the redaction of the manuscript.

\section*{Competing financial interests}
The authors declare no competing financial interests.

\end{document}